\documentclass[10pt,journal,final,finalsubmission,twocolumn]{IEEEtran}
\usepackage{epsfig}
\usepackage{url}
\usepackage{amsmath, amsthm, amssymb}

\newtheorem{example}{Example}
\newtheorem{discussion}{Discussion}

\usepackage{algorithm}
\usepackage{algorithmic}
\usepackage{multirow}
\usepackage{amstext}
\usepackage{blkarray}

\usepackage{graphicx}
\usepackage{subfigure}

\usepackage{tabularx}
\usepackage{threeparttable}
\usepackage{url}
\usepackage{color}

\newcommand{\warn}[1]{}
\newcommand{\nop}[1]{}

\begin{document}

%\conferenceinfo{CCS'09,} {November 9--13, 2009, Chicago, Illinois,
%USA.} \CopyrightYear{2009} \crdata{978-1-60558-352-5/09/11}

\title{Bandwidth-Aware Scheduling with SDN in Hadoop: A New Trend for Big Data}
\newcommand{\superast}{\raisebox{9pt}{$\ast$}}%
\newcommand{\superdagger}{\raisebox{9pt}{$\dagger$}}
\newcommand{\superddagger}{\raisebox{9pt}{$\ddagger$}}
\newcommand{\superS}{\raisebox{9pt}{$\S$}}
\newcommand{\superP}{\raisebox{9pt}{$\P$}}

\author{\IEEEauthorblockN{Peng Qin, Bin Dai, Benxiong Huang} and Guan Xu\\
\IEEEauthorblockA{Department of Electronics and Information Engineering \\
Huazhong University of Science and Technology, Wuhan, China\\}
%\IEEEauthorblockA{\IEEEauthorrefmark{2}School of Information Technology, Deakin University, Australia\\}
Email: \{qinpeng, daibin, huangbx\}@hust.edu.cn, guanxu86@gmail.com}

%\author{\IEEEauthorblockN{Author 1, Author 2, Author 3 and Author 4}}

\maketitle

\begin{abstract}

Software Defined Networking (SDN) is a revolutionary network architecture that separates out network control functions from the underlying equipment and is an increasingly trend to help enterprises build more manageable data centers where big data processing emerges as an important part of applications. To concurrently process large-scale data, MapReduce with an open source implementation named Hadoop is proposed. In practical Hadoop systems one kind of issue that vitally impacts the overall performance is know as the NP-complete minimum make span\footnote{Make span means the time between job's start time and job's finish time.} problem. One main solution is to assign tasks on data local nodes to avoid link occupation since network bandwidth is a scarce resource. Many methodologies for enhancing data locality are proposed such as the HDS~\cite{Tom::HadoopTheDefiniGuide::2012} and state-of-the-art scheduler BAR~\cite{Jiahui::BarAnEffiDataLocalDrivTaskScheAlgoForCloudComp::2011}. However, all of them either ignore allocating tasks in a global view or disregard available bandwidth as the basis for scheduling. In this paper we propose a heuristic bandwidth-aware task scheduler BASS to combine Hadoop with SDN. It is not only able to guarantee data locality in a global view but also can efficiently assign tasks in an optimized way. Both examples and experiments demonstrate that BASS has the best performance in terms of job completion time. To our knowledge, BASS is the first to exploit talent of SDN for big data processing and we believe it points out a new trend for large-scale data processing.

\end{abstract}


\vspace{-0.1cm}%If you provide a negative argument, it will add a negative space, thus removing some white space
\section{Introduction}
\label{sec::introduction}
\vspace{-0.1cm}

Software Defined Networking (SDN)\footnote{https://www.opennetworking.org/.} is a revolutionary network architecture that separates out network control functions from the underlying equipment and deploys them centrally on the controller, where OpenFlow is the standard interface. With SDN, applications can treat the network as a logical entity which makes enterprises and carriers gain unprecedented programmability, automation and network control. In addition, SDN also provides a set of APIs to simplify the implementation of common network services such as routing, multicast, security, access control, bandwidth management, quality of service (QoS) and storage optimization etc~\cite{Siamak::CloudCompNetChallAndOpporForInnova::2013}.

As a result, SDN creates amounts of opportunities to help enterprises build more deterministic, innovative, manageable and high scalable data centers that extend beyond private enterprise networks to public IT resources, which makes implementing OpenFlow based SDN data centers become a new developing trend nowadays.

%while still guaranteeing higher network efficiency and agility to carriers focusing on improving their services' profitability by provisioning more services with fewer, better optimized resources. ~\cite{Siamak::CloudCompNetChallAndOpporForInnova::2013}.

\begin{figure}[htb]
\begin{center}
\epsfig{file =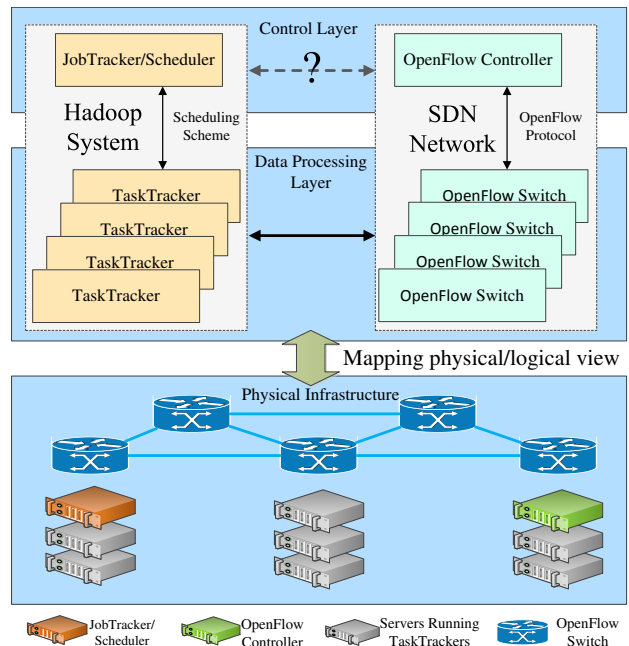,width=0.48\textwidth}
\vspace{-4mm}
\caption{\label{fig::SDNBigDataProcessArchitecture} Architecture of Hadoop Big Data Processing with SDN}
\end{center}
\vspace{-4mm}
\end{figure}

At the same time, big data processing emerges as an important part of applications in such kind of data centers, where they not only handle but also generate amounts of data everyday. To concurrently process large-scale data with high efficiency, MapReduce with an open source implementation named Hadoop\footnote{http://lucene.apache.org/.} is proposed. It is an increasingly common computing system used by Yahoo!, Amazon and Facebook etc. The logical view of Hadoop system is shown on the upper left of Fig.~\ref{fig::SDNBigDataProcessArchitecture} while the physical view is shown at the bottom.

%For Hadoop system, the Scheduler/JobTracker divides a job into several tasks processed by TaskTrackers and for SDN, the Controller maintains a global view of the network with central management capability.

%One important feature is that it introduces the capability of network monitoring which provisions us with the ability to obtain real-time link bandwidth.

In practical Hadoop systems one kind of issue that vitally impacts the overall performance is know as the NP-complete~\cite{MichaelFischer::AssignTaskForEffiInHadoop::2010} minimum make span problem~\cite{KenBirman::ToAClouCompuReseaAgen::2009}~\cite{Edward::OpenSourGridTechForWebScalComp::2009}, which exploits to find solutions on how to minimize the job completion time. Since network bandwidth is a scarce resource, assigning tasks on data local nodes is important for avoiding link occupation and then shortening the make span.

%In practical Hadoop systems network bandwidth is a scarce resource and it has a vital impact on system performance. Therefore, assigning tasks on data local nodes is important for reducing the bandwidth occupation which can shorten the overall job completion time. However, when Hadoop cluster is shared between users, in which case data locality cannot be guaranteed, data movement becomes a must. How to minimize the job completion time is known as the minimum make span\footnote{Make span means the time between job's start time and job's finish time.} scheduling problem~\cite{KenBirman::ToAClouCompuReseaAgen::2009}~\cite{Edward::OpenSourGridTechForWebScalComp::2009} and it is NP-complete~\cite{MichaelFischer::AssignTaskForEffiInHadoop::2010}.

Many methodologies for enhancing data locality are proposed such as the Hadoop Default Scheduler (HDS)~\cite{Tom::HadoopTheDefiniGuide::2012} and state-of-the-art BAlance-Reduce scheduler (BAR)~\cite{Jiahui::BarAnEffiDataLocalDrivTaskScheAlgoForCloudComp::2011}. However, all of them either ignore allocating tasks in a global view or disregard available bandwidth as the basis for scheduling which results in losing optimized opportunities for task assignment.

As the new trend for big data processing evolves with SDN, one question to handle the minimum make span issue is naturally asked: \textbf{Can we combine the bandwidth control capability of SDN with Hadoop system to exploit an optimized
task scheduling solution that has high efficiency and agility in terms of job completion time for big data processing?} (shown by the question mark on top of Fig.~\ref{fig::SDNBigDataProcessArchitecture})

%some gaps and concerns emerge: What are the potential solutions using the existing technologies for the implement of big data processing networks to comply with the development of virtual networks? Is there any room to utilize innovative paradigms such as the Software Defined Networking (SDN) to address networking challenges? How to guarantee data security of the whole system in a more shared environment and how to exploit new solutions to provisions the big data processing with more efficiency, agility and control abilities~\cite{Siamak::CloudCompNetChallAndOpporForInnova::2013}?

In this paper we propose a bandwidth-aware task scheduler BASS (Bandwidth-Aware Scheduling with Sdn in hadoop) to combine Hadoop with SDN. It first utilizes SDN to manage the bandwidth and allocates it in a Time Slot (TS) manner, then BASS decides whether to assign a task locally or remotely depending on the completion time. It is not only able to guarantee data locality in a global view but also can efficiently assign tasks in an optimized way. The most important is that BASS takes a full consideration of the scarce network bandwidth from OpenFlow controller and regards it as a vital parameter for task scheduling. To our knowledge, BASS is the first to exploit talent of SDN for big data processing in Hadoop. Both examples and real world experiments demonstrate that BASS outperforms all previous related algorithms including BAR which represents state-of-the-art.

\vspace{-0.03cm}
\subsection{Contributions}
\vspace{-0.01cm}

%In this paper, we exploit to utilize the management capability of SDN, take sufficient account of scarce bandwidth and propose a heuristic task scheduler SHIDA which minimize the make span of a job in an optimized way.

The main contributions are summarized as follows.

\begin{itemize}
  \item     We formalize the make span problem and develop a TS scheme for bandwidth allocation.
  \item  	We exploit the capability of SDN and propose a bandwidth-aware task scheduler BASS which outperforms all previous related algorithms.
  \item     We provide \emph{Example~\ref{example::shida_task_assignment}, Example~\ref{example::preshida_scheme}, Example~\ref{example::SDN_QoS_scheme}} and implement extensive real world experiments to demonstrate the effectiveness of BASS.
\end{itemize}

\vspace{-0.03cm}
\subsection{Paper structure}
\vspace{-0.01cm}

The remainder of the paper is structured as follows. In Section \ref{sec::related work} we review some related work. In Section \ref{sec::problem} we formalize the problem of scheduling in Hadoop cluster. In Section \ref{sec::SDN_combination_with_Hadoop} we propose the SDN based bandwidth-aware scheduler BASS and present detailed examples for illustration. In Section \ref{sec::experiment_result} we describe experiment details. Section \ref{sec::conclusions} concludes the paper and provides future expectations.

\vspace{-0.1cm}
\section{Related work}
\label{sec::related work}
\vspace{-0.1cm}

A broad class of prior literatures ranging from big data processing to new emerging SDN is related to our work.

The Hadoop default scheduler searches for data local tasks greedily and assigns them to idle nodes~\cite{Matei::JobScheForMulUserMapReduClust::EECS-2009-55} which, however, results in an increased job completion time. Matei et al.~\cite{Matei::DelayScheduling::2010} propose delay scheduling to address the conflict between data locality and fairness. However, the introduced delays may lead to under-utilization and instability. Jian Tan et al.~\cite{JianTan::CouplTaskProgrForMapReduResouAwarSchedu::2013} find that for current schedulers, map tasks and reduce tasks are not jointly optimized, which may cause job starvation and unfavorable data locality. To mitigate this problem they propose a coupling scheduler to combine map and reduce tasks. Since Hadoop assumes that all cluster nodes are dedicated to a single user, it fails to guarantee high performance in shared environment. To address this issue, Sangwon et al.~\cite{SangwonSeo::HPMRprefetchPreshuffling::2009} propose a prefetching and pre-shuffling scheme. However, the bandwidth occupation for transferring data block cannot be significantly reduced.

Jiahui et al.~\cite{Jiahui::BarAnEffiDataLocalDrivTaskScheAlgoForCloudComp::2011} propose the BAR scheduler to globally reduce the job completion time which is the most related to BASS. It is based on prior work~\cite{MichaelFischer::AssignTaskForEffiInHadoop::2010} proposed by Michael et al. to assign tasks efficiently in Hadoop. In~\cite{MichaelFischer::AssignTaskForEffiInHadoop::2010} they investigate task assignment in Hadoop and give an idealized Hadoop model to evaluate the cost of task assignments. It is shown that task assignment is a NP-complete problem which, however, can only be found a near optimal solution with high computation complexity. To address this issue BAR first produces an initial task allocation, then the job completion time can be gradually reduced by tuning the initial task allocation. However, in some cases, such as \emph{Discussion~\ref{discussion::shida_bar_native_comparison}} shows in Section~\ref{sec::SDN_combination_with_Hadoop}, BAR cannot efficiently reduce the job completion time while BASS can reduce it from 39s to 35s.

Independently, SDN originating from Clean Slate by University of Stanford is a new network architecture that separates out network control functions from the underlying equipment. The underlying switches perform only simple data forwarding. The leading SDN technology is based on the OpenFlow protocol~\cite{NickMckeown::OpenFlowEnabInnovaInCampNet::2008}, a standard that has been designed for SDN and already being deployed in a variety of networks and networking products~\cite{Hadoop::HadoopScalFlexDataStorAnaly::2010}~\cite{Sandhya::OpenFlowEnabHadooOverLocaWideAreClus::2012}. The most important feature of OpenFlow is its capability of network monitoring and traffic control which gives an alternative solution to speed up the Hadoop big data processing system.

%Sandhya et al.~\cite{Sandhya::HadooAcceInOpenFlowCluster::2012} focus on utilizing this ability provided by OpenFlow and propose to decrease the execution time for a Hadoop job by provisioning the shuffle traffic more of the available bandwidth on a link. However, the data locality is not enhanced in this work, thus performance improvement is limited. 

\vspace{-0.1cm}
\section{Problem formalization}
\label{sec::problem}
\vspace{-0.1cm}

With the capability of network control provisioned by SDN, we can capture the real time network status such as network traffic and bandwidth. We define some notations as follows~\cite{ZhenhuaGuo::ImproMapReduPerfoInHeterNetEnvirAndResouUtili::2012}:

$TK_i$ denotes a task $i$ within a Hadoop job; $ND_j$ denotes a node $j$ in the Hadoop cluster; $SZ_i$ denotes the size of input split data for $TK_i$ when it is assigned to $ND_j$; $TM_{i,j}$ denotes the data movement time of $TK_i$ from data source $ND_{dataSrc}$ to $ND_j$; $TP_{i,j}$ denotes the time of task computation; $TE_{i,j}$ denotes the time for task execution; $\Upsilon I_{j}$ denotes the time when $ND_j$ becomes idle; $\Upsilon C_{i,j}$ denotes the completion time of $TK_i$; $BW_{j,k}$ denotes the bandwidth between $ND_j$ and $ND_k$ while $BW_{rl}$ denotes the real time available bandwidth of a link. Based on above symbols we obtain Eq.(\ref{eqn::TimeMovment}) to Eq.(\ref{eqn::TimeCompletion}).

\begin{equation}
\label{eqn::TimeMovment}
TM_{i,j}=SZ_i/BW_{dataSrc,j}
\end{equation}

\begin{equation}
\label{eqn::TimeExecution}
TE_{i,j}=TP_{i,j}+TM_{i,j}
\end{equation}

\begin{equation}
\label{eqn::TimeCompletion}
\Upsilon C_{i,j}=TE_{i,j}+\Upsilon I_{j}
\end{equation}

For a map or reduce task $TK_i$, the Objective Function (shown in Eq.(\ref{eqn::ObjecFuncChooseNode})) is to find an available node that can yield the earliest completion time among all $n$ nodes of the cluster.

\begin{equation}
\label{eqn::ObjecFuncChooseNode}
ND_j=argmin_j\Upsilon C_{i,j}
\end{equation}
where $1\leq j\leq n$.

In the aspect of global view for a job, however, the Objective Function (shown in Eq.(\ref{eqn::ObjecFuncJobAspect})) is a little different, where we need to find the slowest map or reduce task $TK_{i'}$ to minimize the completion time of a whole job.

%while on the other hand we need to summate the total execution time of all tasks belonging to a job and try to find a solution to minimize it(Objective Function 2 is shown in Eq.(\ref{eqn::ObjecFuncJobAspectSum})).

\begin{equation}
\label{eqn::ObjecFuncJobAspect}
min\{\Upsilon C_{i',j'}=max\Upsilon C_{i,j}(1\leq i,i'\leq m, 1\leq j,j'\leq n)\}
\end{equation}
where $m$ is the task number of a job and $n$ is the node number of the Hadoop cluster.

%\begin{equation}
%\label{eqn::ObjecFuncJobAspectSum}
%min\Sigma_{i=1}^mTE_{i,j}
%\end{equation}

%Note that the data movement time $TM_{i,j}$ in Eq.(\ref{eqn::TimeMovment}) is an important parameter in our SHIDA scheduler.  

\vspace{-0.1cm}
\section{SDN based bandwidth-aware scheduling in Hadoop for big data processing}
\label{sec::SDN_combination_with_Hadoop}
\vspace{-0.1cm}

\subsection{Time Slot Bandwidth Allocation}

Benefiting from capability of SDN to obtain the real time link bandwidth $BW_{rl}$, we propose a scheme to allocate bandwidth in a Time Slot way. The main principle is described as follows.

%\begin{figure}[htb]
%\begin{center}
%\epsfig{file =TimeSlotScheme.eps,width=0.5\textwidth}
%\vspace{-4mm}
%\caption{\label{fig::TimeSlotScheme} Example: Time slots allocation on a single link}
%\end{center}
%\vspace{-4mm}
%\end{figure}

Before Hadoop task scheduling begins, the occupation time of each link's residue bandwidth is disintegrated into equal time slot, namely, $TS_1,TS_2,...,TS_k,...$, duration of which is a tunable parameter according to practical network scenarios. We use $SL_{rl}$ to denote the residue bandwidth at a certain time slot for a certain link. If the task $TK_i$ has the requirement of data movement through a certain path during $(t_m, t_n)$, the scheduler will assign the corresponding Time Slots to it in advance guaranteeing that the bandwidth of all links on this path from starting slot $TS_m$ ($t_m\in TS_m$) to ending slot $TS_n$ ($t_n\in TS_n$) are reserved for $TK_i$.

%For each task $TK_i$ needing data movement on $Path_{dataSrc,j}$ from $ND_{dataSrc}$ to $ND_j$, the scheduler allocates TSs to it in advance guaranteeing that all the bandwidth on $Path_{dataSrc,j}$ from starting slot $TS_k$ to ending slot $TS_{k+m}$ are occupied by $TK_i$.
%For example, if the occupied bandwidth at time slot $TS_2$ is 60\%, then residue bandwidth $SL_{rl}=40\%$.
%For instance, $SL_{rl}$ at time slots $TS_2$ and $TS_3$ is 100\% while $SL_{rl}$ at $TS_4$ and $TS_5$ is 40\% (1-60\%).

The motivation for proposing TS scheme is as follows. Bandwidth is a scarce resource in practical Hadoop cluster especially when nodes compete drastically. Thus, to sufficiently utilize available bandwidth we argue that always providing tasks requiring data movement with the most residue bandwidth and then taking it back after the occupation is a simple but effective solution in practice. Both \emph{Example~\ref{example::shida_task_assignment}} and real world experiments demonstrate its validity.

\subsection{BASS: Bandwidth-Aware Scheduling with Sdn in hadoop}

The BASS algorithm is illustrated as follows. A submitted job has $m$ tasks and there are $n$ available nodes in a Hadoop cluster. Note that, the number of available nodes $n$ may be less than the total nodes of the cluster especially when Hadoop system is shared by users.

%since in many cases the Hadoop systems for data-intensive applications are usually shared by users with distinct accounts due to practical considerations about cost, system utilization and manageability~\cite{SangwonSeo::HPMRprefetchPreshuffling::2009}. For instance, In Yahoo! a large cluster called Yahoo!Grid is managed by Hadoop-On-Demand (HOD)\footnote{http://hadoop.apache.org/core/docs/r0.18.3/hod.html.}. Each engineer at Yahoo! has a log-on account for the cluster and allocates his or her own virtual nodes on demand which is only a subset of the whole nodes.

\begin{algorithm}[htb]\footnotesize
\caption{BASS algorithm: \textbf{B}andwidth-\textbf{A}ware \textbf{S}cheduling with \textbf{S}dn in hadoop}
\label{alg:SDNbasedHadoopScheduling}
\begin{algorithmic}[1]  %%%%%%%%%%\begin{algorithmic}[1] numbers every line, whereas \begin{algorithmic}[5] numbers every fifth
 \REQUIRE ~~\\%算法的输入参数：Initialization
  Given the submitted job with $m$ tasks $TK_i$ and the $n$ nodes $ND_j$ in a Hadoop cluster.
 \ENSURE ~~\\%算法的主体：Main Body
 \FOR {($i={1,2,...,m}$)}
     \STATE Use capability of SDN to obtain the real-time bandwidth $BW_{rl}$ and the corresponding percentage of residue time slot $SL_{rl}$ for each link;
     \STATE Find $ND_{loc}$ with available idle time $\Upsilon I_{loc}$ for $TK_i$;
     \STATE Find $ND_{minnow}$ with available idle time $\Upsilon I_{minnow}$ for $TK_i$;
     \IF {($ND_{minnow}\equiv ND_{loc}$ $||\Upsilon I_{loc}\leq \Upsilon I_{minnow}$)}
        \STATE Assign $TK_i$ to $ND_{minnow}\equiv ND_{loc}$;
     \ELSIF {($ND_{loc}$ is found $\&\&\Upsilon I_{loc} > \Upsilon I_{minnow}$)}
        \STATE Use Eq.(\ref{eqn::TimeMovment}) to (\ref{eqn::TimeCompletion}) to calculate $\Upsilon C_{i,minnow}$ and the needed bandwidth $BW_{i,minnow}$ guaranteeing that $\Upsilon C_{i,minnow}<\Upsilon C_{i,loc}$;
        \IF {($BW_{i,minnow}\leq BW_{rl}$)}
            \STATE Assign $TK_i$ to remote node $ND_{minnow}$;
            \STATE Assign $SL_{rl}$ of links on path from $ND_{dataSrc}$ to $ND_{minnow}$ and use Eq.(\ref{eqn::TimeMovment}) to calculate the slot number $TK_i$ needs;
            %\STATE Update link bandwidth state with $BW_{rl}-=BW_{i,minnow}$;
        \ELSE
            \STATE Assign $TK_i$ to local node $ND_{loc}$;
        \ENDIF
     \ELSIF {($ND_{loc}$ is not found)}
        \STATE Assign $TK_i$ to remote node $ND_{minnow}$;
        \STATE Assign $SL_{rl}$ of links on path from $ND_{dataSrc}$ to $ND_{minnow}$ and use Eq.(\ref{eqn::TimeMovment}) to calculate the slot number $TK_i$ needs;
     \ENDIF
  \ENDFOR
 \RETURN The assignment for all $m$ tasks.
\end{algorithmic}
\end{algorithm}

\vspace{-0.4cm}
\subsection*{\textbf{Case 1}: Data local node $ND_{loc}$ is found}
\vspace{-0.01cm}

For a task $TK_i$, since bandwidth is a scarce resource in Handoop system we prefer to assign it to a data local node $ND_{loc}$ if there is one. When $ND_{loc}$ is found, its available idle time $\Upsilon I_{loc}$ is recorded. However, $ND_{loc}$ is not always the optimal option especially when there are already too many workloads on $ND_{loc}$ which will let $TK_i$ wait for a nontrivial extra time resulting in a much longer job completion time. In this scenario we take the data movement time $TM_{i,j}$ into account and treat it as a significant parameter for job scheduling. Then we search and find one node $ND_{minnow}$ whose available idle time $\Upsilon I_{minnow}$ is minimum for the current time. $\Upsilon I_{minnow}$ is also recorded for further analysis.

\vspace{-0.4cm}
\subsection*{\textbf{Case 1.1}: Data local node $ND_{loc}$ is optimal}
\vspace{-0.01cm}

If data local node $ND_{loc}$ is just the node $ND_{minnow}$ or its available idle time $\Upsilon I_{loc}$ is no greater than $\Upsilon I_{minnow}$, we assign $TK_i$ to $ND_{loc}\equiv ND_{minnow}$ directly since there is no cost of data movement according to Eq.(\ref{eqn::TimeMovment}).

If data local node $ND_{loc}$ is found but its available idle time $\Upsilon I_{loc}$ is greater than $\Upsilon I_{minnow}$, there will be a tradeoff on whether to assign $TK_i$ to node $ND_{minnow}$ or not, depending on the residue bandwidth $BW_{rl}$ (or $SL_{rl}$ of time slot) of links on path from data source $ND_{dataSrc}$ to $ND_{minnow}$. In this case to make sure that the remote node $ND_{minnow}$ is a better choice than the data local node $ND_{loc}$ for running task $TK_i$, we need to firstly calculate the task completion time $\Upsilon C_{minnow}$ and $\Upsilon C_{loc}$ using Eq.(\ref{eqn::TimeMovment})to Eq.(\ref{eqn::TimeCompletion}).

Subject to the objective that $\Upsilon C_{minnow}$ is smaller than $\Upsilon C_{loc}$ ($\Upsilon C_{minnow}<\Upsilon C_{loc}$), we obtain the corresponding bandwidth needed $BW_{i,minnow}$ for moving data to remote node $ND_{minnow}$. Then we compare $BW_{i,minnow}$ with the real time available bandwidth $BW_{rl}$.

\vspace{-0.4cm}
\subsection*{\textbf{Case 1.2}: Remote node $ND_{minnow}$ is optimal}
\vspace{-0.01cm}

If $BW_{i,minnow}\leq BW_{rl}$, it indicates that the available bandwidth is enough for transferring input data for $TK_i$ with task completion time $\Upsilon C_{minnow}$ earlier than $\Upsilon C_{loc}$. In this case we assign $TK_i$ to remote node $ND_{minnow}$ and assign time slots $SL_{rl}$ on path from data source $ND_{dataSrc}$ to $ND_{minnow}$ according to SDN controller. Note that the TSs on a $link$ that are allocated to task $TK_i$ are determined by the residue TSs of $path$ it belongs to, which are equal to the minimum residue TSs of all its links.

%Take Fig.~\ref{fig::TimeSlotScheme} for instance, for a single link $\ell$ when data is transferred for task $TK_i$ at time slot $TS_1$ the residue bandwidth of its current data path is 100\% which is equal to the idle TS on link $\ell$. However, at time slot $TS_4$ and $TS_5$ the residue bandwidth on the current data path is 60\%, in which scenario even the idle TS on link $\ell$ is still 100\%, the allocated time slot for another task $TK_{i^\prime}$ is only 60\%.

\vspace{-0.4cm}
\subsection*{\textbf{Case 1.3}: $ND_{minnnow}$ is not optimal for limited bandwidth}
\vspace{-0.01cm}

If $BW_{i,minnow}> BW_{rl}$, it indicates that the available bandwidth in not enough for moving input data for $TK_i$ with task completion time $\Upsilon C_{minnow}$ earlier than $\Upsilon C_{loc}$. In this case since the total cost of $ND_{minnow}$ including the data transferring expense is much greater than data local node $ND_{loc}$, there is no need to run task $TK_i$ remotely. Therefore, we assign $TK_i$ to $ND_{loc}$.

\vspace{-0.4cm}
\subsection*{\textbf{Case 2}: Data local node $ND_j$ is not found (locality-starvation)}
\vspace{-0.01cm}

All above cases assume that $ND_{loc}$ can be found and is available. However, Hadoop cluster may be shared by different users and each of whom is only authorized to use a subset of the whole nodes. Thus the input split data has high probability of not being stored in any of these nodes. Therefore, $ND_{loc}$ may not be found in this scenario which we call \emph{locality-starvation} in this paper. To deal with it, \emph{Algorithm \ref{alg:SDNbasedHadoopScheduling}} proposes a similarly solution like the case $BW_{i,minnow}\leq BW_{rl}$ where remote node $ND_{minnow}$ is the optimal.

In this case we assign $TK_i$ to remote node $ND_{minnow}$ and assign time slots $SL_{rl}$ on path from data source $ND_{dataSrc}$ to $ND_{minnow}$ according to SDN controller. The TSs on a $link$ that are allocated to task $TK_i$ are also determined by the residue TSs of $path$ it belongs to, which are equal to the minimum residue TSs of all its links.

All the $m$ tasks are scheduled via above process until an allocation result is obtained. BASS algorithm is an optimized scheduling scheme compared with the HDS and BAR schedulers for the follows three reasons.

\begin{itemize}
  \item     Firstly, BASS maintains the priority of data locality with no cost of transferring data among nodes.
  \item     Secondly, BASS utilizes SDN's bandwidth management capability to assign link bandwidth for a remote node $ND_{minnow}$ making sure that the task completion time $\Upsilon C_{minnow}$ is earlier than that of a local node $ND_{loc}$.
  \item     Last but not least, BASS utilizes a TS scheme to allocate link bandwidth in a temporal dimensionality, which is a simple but effective solution in practice.

%      As is known that bandwidth allocation for all links is a NP-Hard problem where one can not always find an optimal solution. However, we believe that the TS scheme not only simplifies this issue but also takes a full consideration about the capability of a single link and always assigns each with the most bandwidth which guarantees the earliest task completion time $\Upsilon C_{i,minnow}$.
\end{itemize}

\begin{figure}[htb]
\begin{center}
\epsfig{file =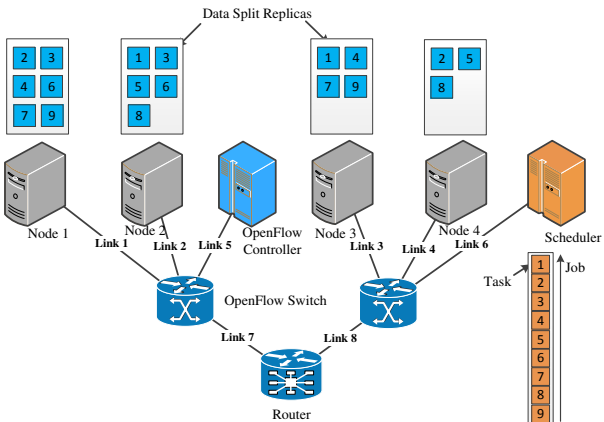,width=0.46\textwidth}
\vspace{-4mm}
\caption{\label{fig::TopologyHadoopSDN} Topology of a Hadoop cluster centrally controlled by SDN}
\end{center}
\vspace{-4mm}
\end{figure}

%\begin{figure*}[htb]
%\begin{center}
%\begin{minipage}{90mm}
%\centering
%\epsfig{file=SHIDAtaskAllocation.eps, width=272pt}
%\vspace{-4mm}
%\caption{\label{fig::SHIDAtaskAllocation} Task allocation result by SHIDA algorithm }
%\vspace{-4mm}
%\end{minipage} \hfil
%\begin{minipage}{90mm}
%\centering
%\epsfig{file=NativeTaskAllocation.eps, width=272pt}
%\vspace{-4mm}
%\caption{\label{fig::NativeTaskAllocation} Task allocation result by Hadoop native scheduler }
%\vspace{-4mm}
%\end{minipage} \hfil
%\end{center}
%\end{figure*}

To explain this algorithm clearly we give the following example.

\begin{example}
    \label{example::shida_task_assignment}
    As is shown in Fig.~\ref{fig::TopologyHadoopSDN}, an OpenFlow controlled Hadoop cluster is composed of 4 task nodes, namely, $Node 1, Node 2, Node 3, Node 4$, an OpenFlow Controller and a Master Node/Scheduler. There are 2 OpenFlow switches, a router and 8 links, namely, $Link 1, Link 2, ..., Link 8$ connecting all nodes. A job with 9 tasks $TK_i~(i=1,...,9)$ is scheduled by the master node. Each input split data has 2 replicas located in 2 different nodes, respectively.

    Assuming that size of each data block is 64MB and each link bandwidth is 100Mbps, if the available bandwidth percentage $SL_{rl}$ is 100\%, then data movement time $TM_{i,j}$ calculated by Eq.(\ref{eqn::TimeMovment}) is 5.12s. Here we choose 5s for simplification. We set each time slot $TS_k$ to be 1s in this paper thus one data block movement occupies 5 time slots. Since each node is homogeneous, task computation time $TP_{i,j}$ is equal and we use 9s for illustration in this example.

%    \begin{figure}[htb]
%    \begin{center}
%    \epsfig{file =SHIDAtaskAllocation.eps,width=0.5\textwidth}
%    \vspace{-4mm}
%    \caption{\label{fig::SHIDAtaskAllocation} Task allocation result by SHIDA algorithm}
%    \end{center}
%    \vspace{-4mm}
%    \end{figure}

    \begin{figure*}[htb]
    \begin{center}
    \epsfig{file =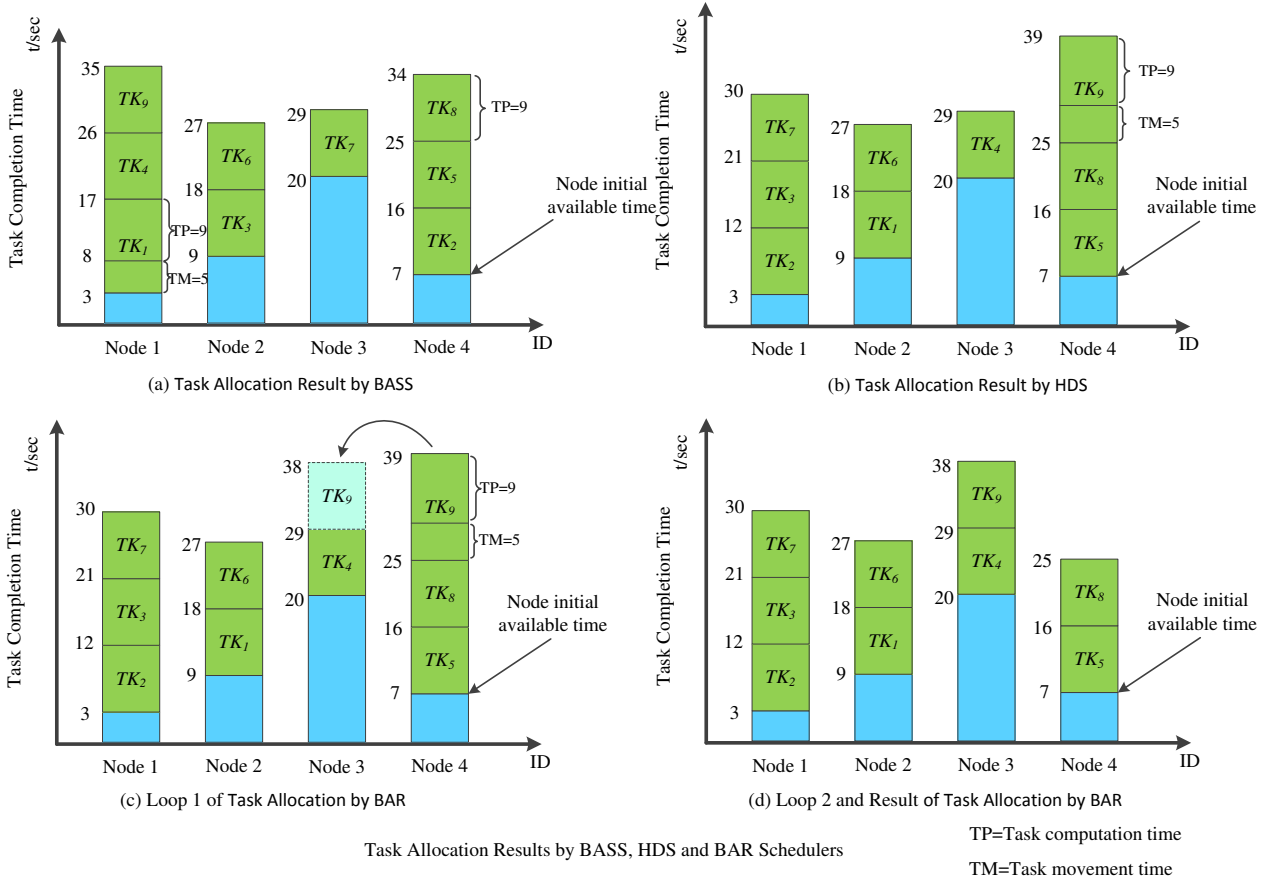,width=0.95\textwidth}
    \vspace{-4mm}
    \caption{\label{fig::AllTaskAllocation} Task allocation results by BASS, HDS and BAR Schedulers}
    \end{center}
    \vspace{-4mm}
    \end{figure*}

    After scheduling starts, BASS allocates tasks $TK_i~(i=1,...,9)$ sequentially as Algorithm \ref{alg:SDNbasedHadoopScheduling} describes and Fig.~\ref{fig::AllTaskAllocation}(a) shows the allocation result. Take the first task $TK_1$ for instance, the available idle time (initial loads) of 4 servers are $\Upsilon I_{1,1}=3s$, $\Upsilon I_{1,2}=9s$, $\Upsilon I_{1,3}=20s$, $\Upsilon I_{1,4}=7s$, respectively.

    For $TK_1$, BASS first finds a node $ND_2$ with data locality and records its available idle time $\Upsilon I_{1,2}=9s$; Then it finds another node $ND_{minnow}=ND_1$ whose available idle time $\Upsilon I_1=3s$ is minimum for the current time; Since $\Upsilon I_{2}>\Upsilon I_1$ which means the available idle time of data local node $ND_2$ is greater than a remote node $ND_1$, then it uses Eq.(\ref{eqn::TimeMovment}), Eq.(\ref{eqn::TimeExecution}), Eq.(\ref{eqn::TimeCompletion}) to calculate the task completion time $\Upsilon C_{1,1}$ and $\Upsilon C_{1,2}$ running on the two nodes. By checking that the residue bandwidth of 100Mbps on Link 1 and Link 2 is enough for moving data block 1 with data transferring time $TM_{1,1}=5s$ and $TM_{1,2}\approx 0s$, it confirms that the task completion time $\Upsilon C_{1,1}=TM_{1,1}+TP_{1,1}+\Upsilon I_{1,1}=5s+9s+3s=17s$ running on remote node $ND_1$ is less than that of running on data local node $ND_2$ with $\Upsilon C_{1,2}=TM_{1,2}+TP_{1,2}+\Upsilon I_{1,2}=0s+9s+9s=18s$. Therefore, it allocates $TK_1$ to $ND_1$. BASS allocates other tasks in the same way. The allocation result is shown clearly in Fig.~\ref{fig::AllTaskAllocation}(a) where we can see the job completion time is 35s since the last task $TK_9$ running on $ND_1$ determines the completion time of a whole job with $\Upsilon C_{9,1}=35s$.

%    \begin{figure}[htb]
%    \begin{center}
%    \epsfig{file =ExampTimeSlotLink1.eps,width=0.5\textwidth}
%    \vspace{-4mm}
%    \caption{\label{fig::ExampTimeSlotLink1} Example:time slots occupation on \textbf{Link 1}}
%    \end{center}
%    \vspace{-4mm}
%    \end{figure}

    In this case BASS transfers input split data for $TK_1$ from $ND_2$, thus the residue bandwidth on Link 1 that is 100\% of 100Mbps from 3s to 8s is allocated for data movement. This means the occupied time slots consist of $TS_4, TS_5, TS_6, TS_7, TS_8$. The occupation of time slots on Link 2 is the same.

    Note that, we may also choose $ND_3$ to transfer input split data for $TK_1$. In this case Link 1, Link 7, Link 8 and Link 3 need to allocate time slots for data movement where the occupation is also the same as before.

\end{example}

\begin{discussion}
    \label{discussion::shida_bar_native_comparison}
    To see the efficiency of BASS scheduler we choose HDS and BAR scheduler that stands for state-of-the-art in this domain to assign the same 9 tasks for comparison. Also assuming that data movement time is 5s and task computation time is 9s.

%    \begin{figure}[htb]
%    \begin{center}
%    \epsfig{file =NativeTaskAllocation.eps,width=0.5\textwidth}
%    \vspace{-4mm}
%    \caption{\label{fig::NativeTaskAllocation} Task allocation result by Hadoop native scheduler}
%    \end{center}
%    \vspace{-4mm}
%    \end{figure}

    HDS always chooses a task $TK_i$ for $ND_{j}$ with data locality and assigns $TK_i$ to it. If no data local task is available, HDS scheduler will choose a task randomly for $ND_j$. Take $TK_1$ for instance, replicas of its input split data are stored at $ND_2$ and $ND_3$, after comparing the available idle time of them, HDS knows $\Upsilon I_2=9s< \Upsilon I_3=20s$, thus, it allocates $TK_1$ to node $ND_2$. Similarly, remaining tasks are assigned. Note that, when $ND_4$ is available at 25s, only non-local $TK_9$ is left, then $ND_4$ has to carry it out.  Finally, $ND_1$ executes $TK_2, TK_3, TK_7$; $ND_2$ executes $TK_1, TK_6$; $ND_3$ executes $TK_4$; $ND_4$ executes $TK_5, TK_8, TK_9$. Since $TK_9$ running on $ND_4$ determines the completion time of the whole job, we see that the job completion time is 39s which is 4s later than that of BASS scheduler (shown in Fig.~\ref{fig::AllTaskAllocation}(b)).

%    \begin{figure*}[htb]
%    \begin{center}
%    \epsfig{file =BARtaskAllocation.eps,width=1\textwidth}
%    \vspace{-4mm}
%    \caption{\label{fig::BARtaskAllocation} Task allocation result by BAR (BAlance-Reducer) scheduler }
%    \end{center}
%    \vspace{-4mm}
%    \end{figure*}

    BAR scheduler is based on HDS and it further improves job performance by globally adjusting data locality according to network state and cluster workload. In the first phase, BAR allocates tasks $TK_i (i=1,...,9)$ to nodes $ND_j (j=1,...,4)$ obeying data locality principle with the same result shown in Fig.~\ref{fig::AllTaskAllocation}(b). In the second phase, BAR searches for the task $TK_{lat}$ whose completion time $\Upsilon C_{i,lat}$ is the latest and checks if there is a remote node $ND_{remo}$ with the completion time $\Upsilon C_{i,remo}$ earlier than $\Upsilon C_{i,lat}$. If $ND_{remo}$ is found BAR assigns $TK_i$ to it and repeats this process until no such node is available. Using the same parameters aforementioned and taking the second phase of BAR scheduler for instance. Since the data local assignment is obtained in the first phase, BAR knows that $TK_9$ on $ND_4$ is the latest one with completion time $\Upsilon C_{9,4}=39s$ (shown in Fig.~\ref{fig::AllTaskAllocation}(c)). It then checks if $ND_3$ with available idle time $\Upsilon I_{3}=25s$ can run task $TK_9$ with completion time $\Upsilon C_{9,3}<39s$? Fortunately, $\Upsilon C_{9,3}=TM_{9,3}+TP_{9,3}+\Upsilon I_3=0s+9s+29s=38s$, is smaller than 39s. Therefore, BAR moves $TK_9$ from $ND_4$ to $ND_3$ with job completion time being 38s, as is shown in Fig.~\ref{fig::AllTaskAllocation}(d).

%    \begin{figure}[htb]
%    \begin{center}
%    \epsfig{file =ExampleComparisonHDSBARSHIDA.eps,width=0.5\textwidth}
%    \vspace{-4mm}
%    \caption{\label{fig::ExampleComparisonHDSBARSHIDA} Performance comparison of three schedulers: HDS, BAR, SHIDA }
%    \end{center}
%    \vspace{-4mm}
%    \end{figure}

    \begin{figure}[htb]
    \begin{center}
    \epsfig{file =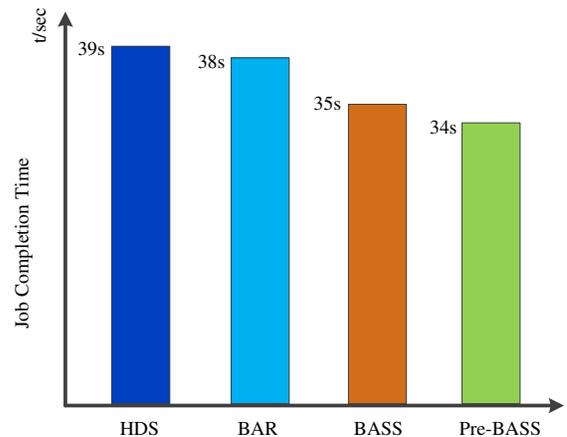,width=0.42\textwidth}
    \vspace{-4mm}
    \caption{\label{fig::ExampleComparisonHDSBARSHIDAPreSHIDA} Performance comparison of schedulers: HDS, BAR, BASS and Pre-BASS }
    \end{center}
    \vspace{-4mm}
    \end{figure}

    From this case we can see clearly that BASS scheduler outperforms BAR and HDS to improve the overall performance in terms of job completion time. A comparison of HDS, BAR and BASS is summarized in Fig.~\ref{fig::ExampleComparisonHDSBARSHIDAPreSHIDA}.

\end{discussion}

\begin{discussion}
    \label{discussion::further_reduce_job_completion_time}
    Can we further reduce the job completion time?

    Sangwon et al. in~\cite{SangwonSeo::HPMRprefetchPreshuffling::2009} propose a prefetching scheme to improve the overall performance under shared environment while retaining compatibility with the native Hadoop scheduler. Inspired by this idea we propose to use a similar prefetching method called Pre-BASS to further reduce the job completion time.

    The main process is described as follows. Firstly, the optimized Pre-BASS scheduler allocates tasks guaranteeing that each one is optimal in terms of its completion time. Then Pre-BASS checks each data-remote task $TK_{remo}$ and let its input split data prefetched/transferred before the available idle time, $\Upsilon I_{remo}$, as early as possible which depends on the real-time residue bandwidth $BW_{rl}$ or $SL_{rl}$. Note that when prefetching a data block, it is always moved from the least loaded node storing the replica to minimize the impact on the overall performance. We give another example to illustrate Pre-BASS scheme.

%    \begin{figure}[htb]
%    \begin{center}
%    \epsfig{file =DiscussioPreSHIDA.eps,width=0.42\textwidth}
%    \vspace{-4mm}
%    \caption{\label{fig::DiscussioPreSHIDA} Further performance melioration of SHIDA }
%    \end{center}
%    \vspace{-4mm}
%    \end{figure}

    \begin{example}
    \label{example::preshida_scheme}
    Thinking about the first task $TK_1$ of BASS running on remote node $ND_1$ (as Fig.~\ref{fig::AllTaskAllocation}(a) shows) in \emph{Example~\ref{example::shida_task_assignment}}. The data movement starts at 3s and it occupies five time slots, namely $TS_4, TS_5, TS_6, TS_7, TS_8$. If we utilize the prefetching scheme here to let this task prefetch its input data at 0s and occupy time slots $TS_1, TS_2, TS_3, TS_4, TS_5$, then the completion time of all tasks on this node will be reduced from 35s to 32s. In this way the last finished task will not be $TK_9$ but $TK_8$ and the job completion time will not be 35s but 34s which is a further performance improvement in the global view (see the right side of Fig.~\ref{fig::ExampleComparisonHDSBARSHIDAPreSHIDA} for performance comparison).
    \end{example}

\end{discussion}

\begin{discussion}
    \label{discussion::further_make_use_of_SDN}
    How can we make the utmost of SDN?

    One important feature of SDN/OpenFlow is its simple QoS scheme via a queuing mechanism. Recall the huge volume of shuffling traffic which consumes large amounts of bandwidth. If Hadoop traffic especially the shuffling traffic is provided with higher priority utilizing the QoS capability of OpenFlow~\cite{Sandhya::HadooAcceInOpenFlowCluster::2012}, we believe that the overall performance of Hadoop system in terms of job completion time will be further reduced. We also give an example to illustrate it.

    \begin{example}
    \label{example::SDN_QoS_scheme}
    Take the topology of SDN controlled Hadoop cluster in Fig.~\ref{fig::TopologyHadoopSDN} for instance. We first set the maximum rate of both OpenFlow switches to be 150Mbps and setup three queues: $Q_1$ with 100Mbps, $Q_2$ with 40Mbps, $Q_3$ with 10Mbps. Then new flow entries are added to direct shuffling traffic to $Q_1$ that has higher link bandwidth and to direct background traffic to $Q_3$ to limit its impact on Hadoop task. The rest of traffic occupy $Q_2$. This simple scheme outperforms that of putting all traffic in the same queue with maximum rate of 150Mbps which is the default scheme.
    \end{example}

\end{discussion}

\vspace{-0.1cm}
\section{Experiments for performance evaluation}
\label{sec::experiment_result}
\vspace{-0.1cm}

In this section we present real world experiments to investigate the effectiveness of BASS. For comparison, two prior most-related schedulers HDS and BAR described in Section \ref{sec::SDN_combination_with_Hadoop} are also implemented.

\subsection{Experiment Setup}

In this experiment the Hadoop cluster with OpenFlow switches is shown is Fig.~\ref{fig::TopologyHadoopSDN}. The cluster consisting of 6 nodes located in 5 physical systems runs the Hadoop version 1.2.1 connected to 2 Open vSwitch (OVS)\footnote{http://openvswitch.org/.}.

%Open vSwitch is a production quality, multilayer virtual switch licensed under the open source Apache 2.0 license. It is designed to enable massive network automation through programmatic extension, while still supporting standard management interfaces and protocols such as OpenFlow standard.

The number of block replicas is set to be 3. The size of data block is 64MB. The maximum link rate is set to be 100Mbps which is a tunable parameter in practice.

We utilize the \emph{ProgressRate} scheme which works well in practice to estimate the initial workload and the available idle time $\Upsilon I$ of each node. The \emph{progress rate} of each task is calculated by $ProgressRate=ProgressScore/T$, where \emph{ProgressScore} represents the task progress between 0 and 1; $T$ is the amount of time the task has been running for. The time to complete is then estimated by $\Upsilon I=\frac{(1-ProgressScore)}{ProgressRate}$.

%\begin{equation}
%\label{eqn::progress_rate}
%ProgressRate=ProgressScore/T
%\end{equation}

%\begin{equation}
%\label{eqn::complete_time}
%\Upsilon I=\frac{(1-ProgressScore)}{ProgressRate}
%\end{equation}

%In the following section SHIDA is evaluated for performance in terms of job completion time.

We choose both Wordcount and Sort jobs\footnote{We choose Wordcount and Sort for test because the former consumes more CPU while the later occupies more disk I/O resources.} as our test case and run them for different workload with data size to be 150MB, 300MB, 600MB, 1GB and 5GB, respectively. We repetitively execute a background job to provide each test with initial workload. Each test is run for 20 times and we use the average value for comparison.

\subsection{Experiment Results}

%By comparing with DFS and BAR we evaluate the data locality and job completion time.

%We record the Map phase completion Time (MT), Reduce phase completion Time (RT), Job completion Time (JT) and we calculate the data Local Ratio $LR=\frac{data~local~task~number}{total~task~number}$.

%\begin{equation}
%\label{eqn::JobCompleRatio}
%JR=\frac{job~completion~time~of~XX}{job~completion~ime~of~SHIDA}
%\end{equation}
%
%\begin{equation}
%\label{eqn::DataLocaRatio}
%LR=\frac{data~local~task~number}{total~task~number}
%\end{equation}

\begin{figure}[htb]
\begin{center}
\epsfig{file =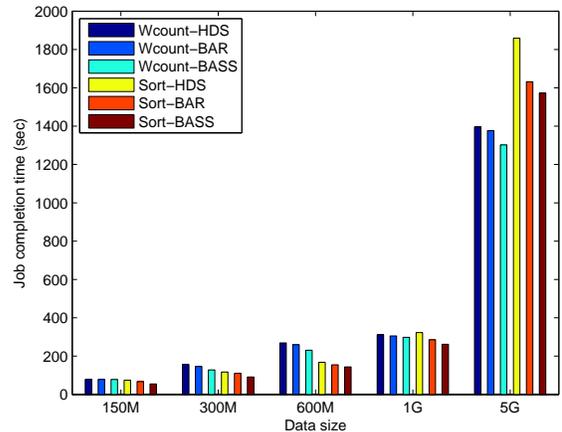,width=0.46\textwidth}
\vspace{-4mm}
\caption{\label{fig::CompletionTimeBoth} Job Completion Time for Both Wordcount and Sort Jobs}
\end{center}
\vspace{-4mm}
\end{figure}

From Table~\ref{tab::dataLocalityJobCompletionTimeWordCount} we see that for Wordcount job, BASS scheduler always finishes with the minimum map and reduce phase completion time, namely the minimum make span compared with BAR and HDS. In some cases data locality ratio (LR) of BASS is low, taking the input data of 600M for example, LR of BASS is 58.3\% while LR of BAR and HDS are 66.7\% and 83.3\%. However, the make span of BASS is only 231s compared with 259s of BAR and 269s of HDS.

The reason is that in this scenario bandwidth resource is sufficient for transferring data and computation resource on data local node $ND_{loc}$ is scarce, thus running $TK_i$ on $ND_{loc}$ is not a good choice anymore. In this sight we argue that link bandwidth and data locality should be taken integratedly into account to achieve the best overall performance in practice.

Sort job in Table~\ref{tab::dataLocalityJobCompletionTimeSort} also demonstrates the validity of BASS scheduler in real world system.

\begin{table*}[htb]
\caption{ Comparison of data locality and job completion time among BASS, BAR and HDS shedulers}
\begin{center}
\subtable[Wordcount jobs]{
\label{tab::dataLocalityJobCompletionTimeWordCount}
\begin{tabular}{|c|c|c|c|c|c|c|c|c|c|c|c|c|c|c|c|}
\hline
\multirow{2}{*}{Data size} &
\multicolumn{4}{c|}{BASS} &
\multicolumn{4}{c|}{BAR} &
\multicolumn{4}{c|}{HDS} \\
\cline{2-13}
  & MT(s) & RT(s) & JT(s) & LR & MT(s) & RT(s) & JT(s) & LR & MT(s) & RT(s) & JT(s) & LR\\
\hline
150M & 58 & 52 & 78  & 33.3\% & 59 & 47 & 78 & 33.3\% & 65 & 53 & 78 & 33.3\% \\
\hline
300M & 79 & 76 & 128 & 66.7\%  & 83 & 105 & 146 & 66.7\% & 89 & 107 & 156 & 50\% \\
\hline
600M & 149 & 192 & 231 & 58.3\% & 183 & 220 & 259 & 66.7\% & 193 & 232 & 269 & 83.3\% \\
\hline
1G & 275 & 216 & 298 & 76.2\% & 290 & 230 & 305 & 76.2\% & 293 & 232 & 311 & 57.1\% \\
\hline
5G & 1190 & 1164 & 1302 & 75.2\%  & 1320 & 1218 & 1377 & 75.2\% & 1347 & 1252 & 1396 & 77.2\% \\
\hline
\end{tabular}
}

\subtable[Sort jobs]{
\label{tab::dataLocalityJobCompletionTimeSort}
\begin{tabular}{|c|c|c|c|c|c|c|c|c|c|c|c|c|c|c|c|}
\hline
\multirow{2}{*}{Data size} &
\multicolumn{4}{c|}{BASS} &
\multicolumn{4}{c|}{BAR} &
\multicolumn{4}{c|}{HDS} \\
\cline{2-13}
  & MT(s) & RT(s) & JT(s) & LR & MT(s) & RT(s) & JT(s) & LR & MT(s) & RT(s) & JT(s) & LR\\
\hline
150M & 24 & 43 & 55 & 50\% & 25 & 49 & 67 & 25\% & 28 & 62 & 74 & 50\% \\
\hline
300M & 26 & 65 & 91 & 66.7\%  & 47 & 98 & 110 & 66.7\% & 54 & 98 & 117 & 50\% \\
\hline
600M & 79 & 123 & 144 & 50\%  & 90 & 135 & 155 & 50\% & 100 & 148 & 168 & 50\% \\
\hline
1G & 147 & 254 & 262 & 56.3\% & 150 & 261 & 285 & 56.3\% & 152 & 297 & 323 & 62.5\% \\
\hline
5G & 640 & 1531 & 1572 & 66.3\% & 660 & 1600 & 1632 & 71\% & 772 & 1730 & 1859 & 63.7\% \\
\hline
\end{tabular}
}
\begin{tablenotes}
        \footnotesize
        \centering
        \item[1] $MT=Map~phase~completion~Time~(sec)$, $RT=Reduce~phase~completion~Time~(sec)$
        \item[2] $JT=Job~completion~Time~(sec)$, $data~Locality~Ratio(LR)=\frac{data~local~task~number}{total~task~number}$
\end{tablenotes}
\end{center}
\end{table*}

%\begin{figure}[htb]
%\begin{center}
%\epsfig{file =CompletionTimeWordcount.eps,width=0.49\textwidth}
%\vspace{-4mm}
%\caption{\label{fig::CompletionTimeWordcount} Job Completion Time for Wordcount}
%\end{center}
%\vspace{-4mm}
%\end{figure}

A clear comparison of above three schedulers is shown in Fig.~\ref{fig::CompletionTimeBoth} for both Wordcount and Sort jobs, where we can see BASS outperforms the other two in terms of job completion time.

%For Wordcount jobs, Fig.~\ref{fig::CompletionTimeBoth}(a) shows that......

%\begin{figure}[htb]
%\begin{center}
%\epsfig{file =CompletionTimeSort.eps,width=0.49\textwidth}
%\vspace{-4mm}
%\caption{\label{fig::CompletionTimeSort} Job Completion Time for Sort}
%\end{center}
%\vspace{-4mm}
%\end{figure}

\vspace{-0.1cm}
\section{Conclusions and Expectations}
\label{sec::conclusions}
\vspace{-0.1cm}

In this paper, we exploit to utilize SDN and take a full account of link bandwidth for improving performance of big data processing. We first formalize the make span problem in Hadoop and develop a Time Slot (TS) scheme for bandwidth allocation. Then we propose the SDN based bandwidth-aware scheduler BASS which can flexibly assign tasks in an optimized way. Last but not least, We provide examples and implement extensive real world experiments to demonstrate the effectiveness of BASS. To our knowledge, BASS is the first to exploit talent of SDN for big data processing in Hadoop cluster.

As the evolvement of SDN with big data processing, for future work we plan to implement BASS in enterprise's data centers composed of practical SDN products such as the OpenFlow switch and we will evaluate BASS's scalability in a much larger network cluster. Furthermore, we believe that BASS points out a new trend for large-scale data processing.

%As the next step, we plan to implement the \emph{Pre-BASS} and \emph{QoS} scheme to further reduce the overall make span. Moreover, we plan to test BASS in a much larger cluster. 

%\newpage

\bibliographystyle{IEEEtran}\scriptsize
\bibliography{Reference}
%Latex 设置字体大小命令由小到大依次为：
%
%\tiny
%\scriptsize
%\footnotesize
%\small
%\normalsize
%\large
%\Large
%\LARGE
%\huge
%\Huge

%\input{AppendixA}
%\input{AppendixB}
%\input{AppendixC}
%\input{AppendixD}

\end{document}